\documentclass[twocolumn,showpacs,preprintnumbers,aps]{revtex4}
\usepackage{graphicx}
\usepackage{amsmath}
\usepackage{euscript}

\renewcommand{\d}{\mathrm{d}}

\begin{document}

\title{An atom interferometer enabled by spontaneous decay}
\author{R.A. Cornelussen, R.J.C. Spreeuw, and H.B. van Linden van den Heuvell}
\affiliation{Van der Waals - Zeeman Institute, University of Amsterdam, \\
Valckenierstraat 65, 1018 XE  Amsterdam, The Netherlands \\
e-mail: heuvell@science.uva.nl}

\date{\today}

\begin{abstract} We investigate the question whether Michelson type interferometry is possible if
the role of the beam splitter is played by a spontaneous process. This question arises from an inspection of
trajectories of atoms bouncing inelastically from an evanescent-wave (EW) mirror. Each final velocity can be reached
via two possible paths, with a {\it spontaneous} Raman transition occurring either during the ingoing or the outgoing
part of the trajectory. At first sight, one might expect that the spontaneous character of the Raman transfer would
destroy the coherence and thus the interference. We investigated this problem by numerically solving the Schr\"odinger
equation and applying a Monte-Carlo wave-function approach. We find interference fringes in velocity space, even when
random photon recoils are taken into account.\end{abstract}


\maketitle

\section{Introduction}

Spontaneous emission is generally considered a detrimental effect in atom interferometers. The associated random recoil
reduces or even completely destroys the visibility of the interference fringes. In this paper we describe an atom
interferometer where the beam splitter works by means of a spontaneous Raman transition. Our central question will be
whether one can observe interference in such an interferometer which is enabled by a spontaneous process and where
decoherence is built into the beam splitting process from the start.

The role of spontaneous emission in (atom) interferometers has long been connected to the concept of which-way
information, ultimately tracing back to Heisenberg's uncertainty principle \cite{Heisenberg27}. Feynman
\cite{FeynmanLectures} discussed a {\it gedankenexperiment} using a Heisenberg microscope to determine which slit was
taken by a particle in a Young's two-slit interferometer. The scattered photon, needed to determine the position,
spoils the interference pattern due to its associated recoil. Similarly, spontaneous emission in an (atom)
interferometer may provide position information on the atom, while at the same time randomizing the momentum and
spoiling interference \cite{ChaHamPri95,HacHorArn04}. On the other hand, resonance fluorescence, including that from
spontaneous Raman transitions, is as coherent as the incident light in the limit of low saturation \cite{Loudon83,
CCT}. Experimental demonstrations have been given by Eichmann {\it et al.} \cite{EicBerRai93} and Cline {\it et al.}
\cite{CliMilHei94}. An experiment by D\"urr {\it et al.} \cite{DurNonRem98} has made it clear that availability of
which-way information should be conceptually separated from the presence of random recoils. Which-way information can
be obtained without random recoils, nevertheless leading to loss of interference. Here we show that random recoils do
not lead to loss of interference as long as the spontaneously emitted photons do not yield which-way information.

\begin{figure}[t]
\includegraphics[width=65mm]{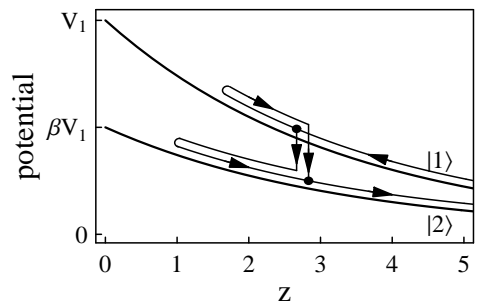}
\caption{Atoms approaching an evanescent-wave mirror in state $|1\rangle$ can undergo a spontaneous transition to state
$|2\rangle$, either on the ingoing or on the outgoing part of the trajectory. These two paths possibly interfere, with
a phase difference depending on the transition point $z$.} \label{fig1}
\end{figure}

Our proposed interferometer is based on cold atoms that reflect from an evescent-wave (EW) mirror. Our model atom is a
typical alkali atom with two hyperfine ground states. During the reflection from the EW mirror the atoms can make a
spontaneous Raman transition to the other hyperfine state. When the repulsive potential experienced by the final state
is lower, the atoms lose kinetic energy and hence bounce inelastically from the potential \cite{HelRolPhi,OvcSodGri95}.
This Sisyphus process has been investigated previously \cite{DesArnDal96} and the resulting final velocity distribution
has been shown to be a caustic \cite{WolVoiHeu01}, reminiscent of the rainbow. Furthermore it is used in several
experiments as the loading process of (low-dimensional) traps for atoms \cite{OvcManGri97,GauHarMly98,CorAmeHeu02}.
Cognet {\it et al.} \cite{CogSavAsp98} observed the analog of St\"uckelberg oscillations in the transverse velocity
distribution of atoms that reflect elastically from a corrugated EW potential. However, no stochastic or incoherent
processes were involved.

In our case the final velocity of an atom depends on the position where it made the Raman transfer \cite{WolVoiHeu01}.
Looking at the trajectories in detail we see that each final velocity can be reached by two trajectories, as is shown
in Fig. \ref{fig1}. An atom can be transferred to the second state on the ingoing or the outgoing part of its
trajectory. Interference will manifest itself in the velocity distribution of the reflected atoms, since the outgoing
velocity depends on the transition point $z$. Note that the beam splitter, its role being played by a spontaneous Raman
transition, is highly non-unitary: atoms are only transferred from state $|1\rangle$ to state $|2\rangle$ and not vice
versa.

This paper is structured as follows. We first present a semi-classical picture, and use it to make qualitative
predictions about the behavior of the interference effects, if present. The question whether interference is possible
will be answered by solving the Schr\"odinger equation in two different ways. The first approach will employ stationary
analytical solutions of the time-independent Schr\"odinger equation, but is limited to monochromatic wave functions.
The second approach will propagate a wave packet by numerically integrating the time-dependent Schr\"odinger equation
with random quantum jumps describing the Raman transitions. The last section deals with experimental considerations.

\section{Semi-classical description}

\begin{figure}[t]
\includegraphics[width=\linewidth]{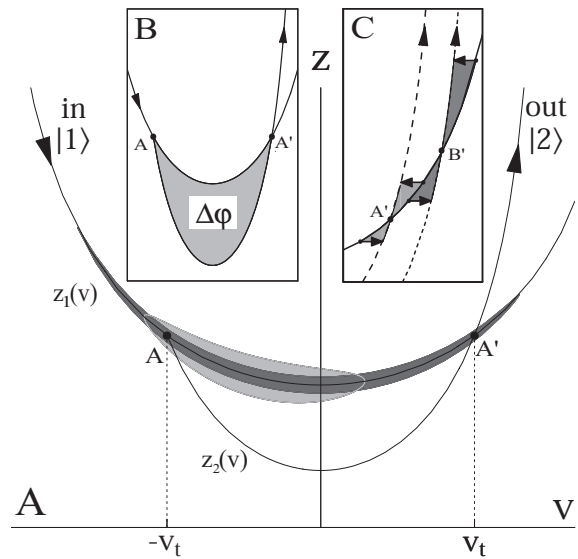}
\caption{Phase-space trajectories of atoms being repelled by an evanescent potential. Atoms initially in state
$|1\rangle$ can be transferred to state $|2\rangle$ and continue on a different path through phase space. (a) Depending
on the initial shape of the wave packet (e.g. the light and dark grey areas) which-way information can be obtained or
not. (b) The accumulated phase difference between these two paths, indicated by the enclosed gray area, may give rise
to interference effects. (c) A momentum kick due to the spontaneous recoil gives rise to an extra phase contribution.}
\label{fig2}
\end{figure}

The analysis will be for a two-level atom, and only the motion of the atom in the direction along the EW-potential
gradient will be considered. The calculations throughout this paper will assume a low saturation parameter, so that
depletion of the initial state can be neglected. An atom in state $|1\rangle$ that reflects from an EW mirror
experiences a potential $V_1\exp(-2\kappa z)$, with $\kappa ^{-1}$ the decay length of the EW field. The atom's
trajectory through phase-space is given by $z_1(v)=(-1/2\kappa)\ln\left[(m/2V_1)(v_{\rm i}^2-v^2)\right]$, with $v_{\rm
i}$ the velocity with which the atom enters the potential, see Fig. \ref{fig2}. After a transition to state $|2\rangle$
in point $A$ or $A'$ the atom experiences a potential $\beta V_1\exp(-2\kappa z)$, with $\beta<1$ the factor by which
the potential energy is reduced after the Raman transition. The atom continues its way through phase space on a new
trajectory $z_2(v)$, given by $z_2(v)=(-1/2\kappa)\ln\left[(m/2\beta V_1)(v_{\rm f}^2-v^2)\right]$, with $v_{\rm f}$
the asymptotic velocity with which the atom leaves the potential. The final velocity can have any value between $v_{\rm
f}=\sqrt{\beta}v_{\rm i}$, which corresponds to a transfer in the turning point, and $v_{\rm f}=v_{\rm i}$, which
corresponds to a transfer outside the EW potential. The final velocity depends on the atom's velocity $v_{\rm t}$ at
the moment of the transfer to state $|2\rangle$. This dependence is given by $v_{\rm f}^2=v_{\rm t}^2+\beta(v_{\rm i}^2
-v_{\rm t}^2)$, from which it is clear that two values $\pm v_{\rm t}$ lead to the same final velocity $v_{\rm f}$. The
two transitions lead to two possible trajectories through phase space, as shown in Fig. \ref{fig2}(a). The phase
difference $\Delta\varphi$ between the two trajectories is given by
\begin{equation}
\Delta\varphi=\frac{m}{\hbar}\int_{-v_{\rm t}}^{v_{\rm t}}\left(z_1(v)-z_2(v)\right)\d v, \label{phasedifference}
\end{equation} This phase difference is proportional to the area between the two curves, indicated in gray in
Fig.~\ref{fig2}(b). From the evaluation of the integral for various parameters we learn that the fringe period
decreases for increasing initial velocities $v_{\rm i}$, for increasing final velocities $v_{\rm f}$, for smaller
$\beta$ and for larger decay length (smaller $\kappa$).

In a semi-classical picture the atom can be treated as a wave packet which is subject to Heisenberg's uncertainty
relation $\Delta z \Delta v_z\geq\hbar/2m$. The distribution of uncertainty between position $z$ and momentum $mv_z$ is
determined by the experimental preparation procedure of the atoms. For a wave packet that initially has a large spread
in momentum, it is not possible to unambiguously determine the phase difference between the two possible paths, since
the wave packet is spread out over several classical trajectories through phase space. This is indicated by the light
grey area in Fig. \ref{fig2}(a). It is, however, possible to determine whether the transfer to state $|2\rangle$ is on
the ingoing or outgoing path, by observing the timing of the spontaneously emitted photon. Therefore, it is not
expected that wave packets with this shape show interference. On the other hand, a minimum uncertainty wave packet with
a narrow initial momentum spread will more closely follow a classical trajectory through phase space. This is indicated
by the dark grey area in Fig. \ref{fig2}(a). The phase difference between the two paths is well defined. The two points
in phase space, A and A', where a transition to the final trajectory is possible are covered simultaneously by the wave
packet. Thus no which-way information can be obtained by observing the time of emission of the spontaneously emitted
photon. The initial trade-off between position and momentum uncertainty in a bandwidth limited wave packet determines
whether interference can be a priori excluded or not.

The random direction of the spontaneously emitted photon can be taken into account in the motion of the atom by a
random momentum jump. This makes the atom propagate on a different trajectory through phase space than it would have
without the random recoil. The momentum changes are indicated by horizontal arrows in the phase-space diagram of Fig.
\ref{fig2}(c). For a single atom, or a collection of distinguishable atoms, the spontaneous recoil could be measured by
detecting the direction in which the photon was emitted. Due to this possibility there will be a set of interference
patterns, one for each recoil direction. By disregarding the information present in the scattered photons, we probe the
incoherent sum of all these interference patterns. The phase difference between the two paths is different with respect
to the recoil free case, and it depends on the direction of the recoil. It is indicated by the gray areas in Fig.
\ref{fig2}(c). In order for the interference to be experimentally observable the difference between the interference
patterns with a certain recoil direction should not be too large. This means that the phase difference between recoil
components in the $\pm z$ directions should be less than $\pi$. For larger final velocities these phase corrections get
larger as is apparent from comparing the areas around the point B' with the areas around point A' in Fig.
\ref{fig2}(c). We thus expect the visibility of the interference to decrease for larger final velocities.

Note that which-way information can neither be retrieved from a measurement of the frequency $\omega$ of the emitted
photon. This frequency is determined by energy conservation: $\hbar\omega=\frac{1}{2}m(v_{\rm i}^2-v_{\rm
f}^2)+\hbar\omega_{\rm EW}-\hbar\Delta_{\rm 12}$, where $\omega_{\rm EW}$ is the frequency of the evanescent photon and
$\hbar\Delta_{\rm 12}$ is the energy difference between states $|1\rangle$ and $|2\rangle$. Because the initial and
final kinetic energies are equal for both trajectories, the frequency of the spontaneously emitted photon is equal for
both interfering paths.


\section{Time-independent approach}

The question whether interference is visible will be answered in this section by considering analytical solutions of
the time-independent Schr\"odinger equation
\begin{equation}
-\frac{\hbar^2}{2m}\frac{\partial^2}{\partial z^2}\psi_1(z)+V_1e^{-2\kappa z}\psi_1(z)=\frac{p_0^2}{2m}\psi_1(z),
\end{equation}
describing stationary states with a total energy $p_0^2/2m$ on the potential $V_1\exp(-2\kappa z)$. It thus describes
particles with momenta $\pm p_0$ in the asymptotic limit of large $z$. This is one of the few examples where the
eigenfunctions of the Schr\"odinger equation are known analytically. The solutions are given by
\begin{equation}
\psi_1(z)=\sqrt{\frac{4p_0}{\pi\hbar\kappa}\sinh\left(\frac{\pi
p_0}{\hbar\kappa}\right)}K_{ip_0/\hbar\kappa}\left(\frac{\sqrt{2mV_1}}{\hbar\kappa} e^{-\kappa z}\right),
\label{BesselKeigenfunction}
\end{equation}
where $K_{\alpha}(\cdot)$ is the Bessel-K function of order $\alpha$ \cite{AbrSte72}. These functions are normalized
such that the asymptotic density is independent of $p_0$. They are also given by \cite{HenCouAsp94} where a different
normalization is used.

This wave function describes the atoms that are incident in state $|1\rangle$. After the spontaneous Raman transition
to state $|2\rangle$, the atoms are described by one of the eigenfunctions $\psi_{2,p}(z)$ with final momentum $p$ in
the potential $V_2\exp(-2\kappa z)$. The final wave function in momentum space is given by the overlap integral
\begin{equation}
 \phi_{k}(p)\propto\int_{-\infty}^{\infty}\psi_1(z)e^{-\kappa z-i k z}\psi_{2,p}^*(z)\d z, \label{timedependentk}
\end{equation}
where the recoil due to the absorbed evanescent photon is taken into account by the factor $\exp(-\kappa z)$ and the
recoil due to the spontaneously emitted photon by the factor $\exp(-i k z)$, with $\hbar k$ the momentum component of
the recoil in the $z$ direction. As already discussed there will be interference patterns $|\phi_{k}(p)|^2$ for every
value $\hbar k$ of the recoil. A measurement that disregards the emitted photon yields the sum of all these possible
interference patterns. In this derivation we will assume an isotropic distribution of the recoil momentum $\hbar k$.
This is an approximation, since the distribution depends on the polarization of the spontaneously emitted photon. We
will come back to this point in section \ref{sec5}. This leads to
\begin{equation}
|\phi(p)|^2=\int_{-k_0}^{+k_0}|\phi_{k}(p)|^2\d k, \label{tindependent}
\end{equation}
with $\hbar k_0$ the total recoil momentum.

\begin{figure}[t]
\includegraphics[width=60mm]{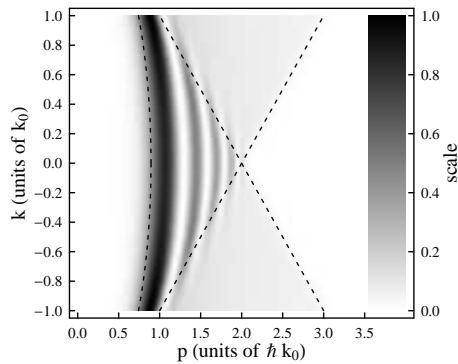}
\caption{The behavior of the interference pattern $|\phi_k(p)|^2$ versus the final momentum $p$, expressed in units of
the photon recoil $\hbar k_0$, for various values of the $z$ component of the photon recoil $\hbar k$ for parameters
$p_0=2\hbar k_0$, $\kappa=k_0/8$, and $\beta=0.2$. The dashed curve indicates the classically lowest reachable
momentum. Interference is visible in the area that can be reached by two trajectories, enclosed by the dashed curve and
straight lines. The triangular regions can only be reached by one trajectory and hence no interference is visible. }
\label{fig3}
\end{figure}

Fig. \ref{fig3} shows the behavior of the momentum distribution for various values of the $z$ component of the photon
recoil $\hbar k$. It is calculated using Eq. \eqref{timedependentk}, with an initial momentum $p_0=2\hbar k_0$, a
potential steepness $\kappa=k_0/8$, and potential reduction $\beta=0.2$. The main features of this figure can be
understood from semiclassical arguments.

The dashed curved and straight lines demarcate three separate classically allowed regions, taking into account the
recoil. The dashed curve on the left indicates the lowest classically reachable momentum, given by
$p=\sqrt{\beta}p_0\sqrt{1-(\hbar k/p_0)^2/(1-\beta)}$. To the right of this curve, we clearly see a region with
interference. In addition we see two triangular regions with no interference, demarcated by two straight lines
described by $p=p_0\pm\hbar k$. The triangular regions can be reached by a single phase space trajectory only. The
upper (lower) triangle corresponds to a trajectory where the Raman transfer took place on the outgoing (incoming)
branch. Only the region on the left of the triangles is reachable by two trajectories and thus shows interference.

The left dashed curve is reached by atoms that scatter a photon near the turning point. Note that in this case the
photon recoil has only a small influence on the final momentum $p$. The final momentum is mainly determined by the
potential energy near the turning point that is converted to kinetic energy. The amount of kinetic energy that can be
added or removed by the photon recoil near the turning point is small because the atomic velocity is small. As a result
we see only a slight curvature as a function of the recoil $k$. For larger values of the initial momentum $p_0$ or the
ratio $\beta=V_2/V_1$ the left curve will become more and more straight.

As expected, the main part of the momentum distribution is in the classically allowed regions. The distributions peak
near the lower classical limit (the dashed curve), resembling the caustic distribution \cite{WolVoiHeu01}. For every
recoil direction interference is visible. Although the region with interference is smaller for larger values of the
recoil, the remaining interference fringes are present at more or less the same final momenta. This indicates that the
spontaneous recoil does not completely wash out the interference. The behavior of the interference does not depend on
the sign of the recoil, because a photon that is emitted on the ingoing part of the trajectory has the same effect on
the momentum distribution as a photon that is emitted in the opposite direction on the outgoing part of the trajectory.

\begin{figure}[t]
\includegraphics[width=\linewidth]{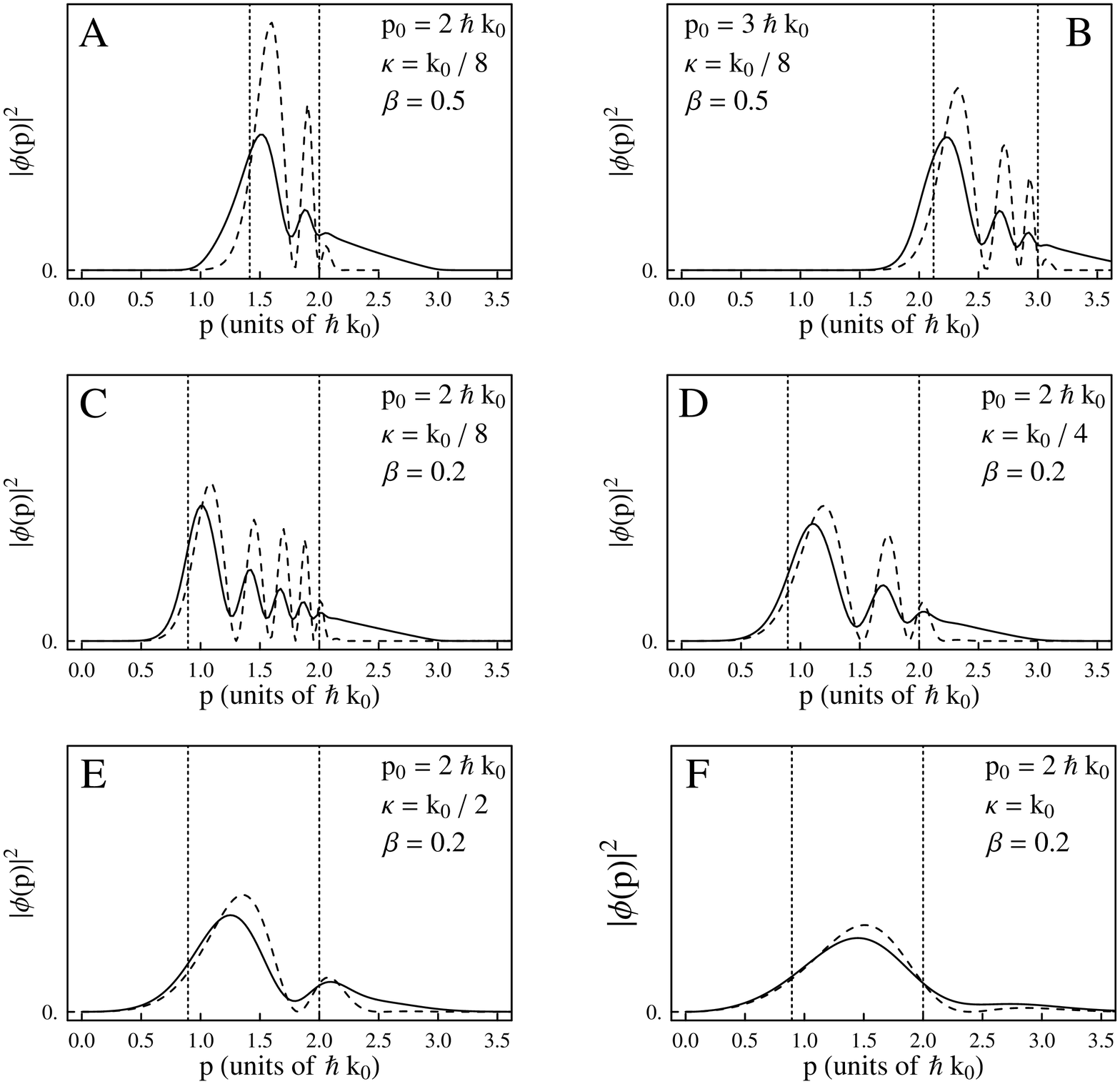}
\caption{Final momentum distributions $|\phi(p)|^2$ calculated for different parameters using the time-independent
method. Solid lines including the effect of a spontaneous recoil, and dashed lines without the effect of recoils,
versus the final momentum $p$ in units of the photon recoil $\hbar k_0$. The dotted lines indicate the classically
allowed region without recoil. Between (a) and (b) the initial momentum $p_0$ is changed, (a) and (c) differ in the
reduction factor $\beta$, (c)-(f) are a sequence for decreasing decay length $\kappa^{-1}$.}\label{fig4}
\end{figure}

Fig. \ref{fig4} shows the final momentum distribution, calculated by Eq. \eqref{tindependent} for various experimental
parameters. The results are compared with an evaluation without a stochastic contribution. Indeed, the averaging over
the spontaneous recoil does not destroy the interference pattern. The small part of the distribution that extends into
the classically forbidden region is an evanescent matter wave. Several of our predictions that were made for the
general case of a coherent interferometer are noticeable in these graphs. Indeed, the fringe spacing decreases for
larger initial and final momenta, for longer decay lengths $\kappa^{-1}$ of the evanescent field, and for smaller
values of $\beta$. Furthermore, as predicted for the case of an incoherent interferometer, the visibility of the graphs
in which the effect of the recoil has been taken into account decreases for larger final momenta.

\section{Time-dependent approach}

In the previous section we have shown that the incoherent nature of the spontaneous Raman transfer does not prevent us
from observing interference. In this section we show that the interference phenomena will also be visible for a wave
packet with a finite momentum spread. In the analysis we closely follow the Monte-Carlo wave-function approach
\cite{DalCasMol92,MolCasDal93}.

We consider the evolution of a diffraction limited wave packet in state $|1\rangle$
\begin{equation}
\psi_1(z,t=0)=\sqrt{\frac{1}{(2\pi)^{1/2}\sigma_{\rm z}}}e^{i k_z z}e^{-\frac{(z-z_0)^2}{4\sigma_{z}^2}}
\end{equation}
with initial height $z_0$, initial width $\sigma_z$, and initial momentum $p_0=\hbar k_z$. It is normalized such that
$\int|\psi(z,0)|^2\d z=1$ and $\int(z-z_0)^2|\psi(z,0)|^2\d z=\sigma_z^2$. The evolution of the wave packet when it
reflects from the evanescent-wave potential with a potential height $V_1$ at $z=0$ is calculated by numerically solving
the time-dependent Schr\"odinger equation
\begin{equation}
i\hbar\frac{\partial}{\partial t}\psi_1(z,t)=-\frac{\hbar^2}{2m}\frac{\partial^2}{\partial
z^2}\psi_1(z,t)+V_1e^{-2\kappa z}\psi_1(z,t)
\end{equation}
using the {\it Quantum kernel} \cite{ThaLie00} package in {\it Mathematica} \cite{mathematica}. This results in a wave
packet $\psi_1(z,t)$ at time t. At a time $\tau$ a spontaneous Raman transition to state $|2\rangle$ occurs, and the
evolution abruptly continues on a potential that is a factor $\beta$ lower. Immediately after the transfer the wave
function in state $|2\rangle$ is described by
\begin{equation}
 \psi_{\tau,k}(z,t=\tau)=\EuScript{N}\psi_1(z,\tau)e^{-\kappa z}e^{-i k z},
\end{equation}
where $\EuScript{N}$ denotes a normalization factor. The two exponents are equal to the exponents in Eq.
\eqref{timedependentk}. The evolution of this wave function can now be continued up to a time $t_{\rm end}$, leading to
a wave function $\psi_{\tau,k}(z,t_{\rm end})$. When $t_{\rm end}$ is large enough the entire wave packet effectively
propagates in free space, so that the momentum distribution remains constant. The Fourier transform
\begin{equation}
 \phi_{\tau,k}(p)=\EuScript{F}\left(\psi_{\tau,k}(z,t_{\rm end})\right)
\end{equation}
of the wave packet at this time is the wave function in momentum space that endured a momentum kick $\hbar k$ at its
transfer time $\tau$. The transfer rate $\Gamma(\tau)$ at a certain time $\tau$ is given by
\begin{equation}
 \Gamma(\tau)\propto\int_0^{\infty}\psi_1^*(z,\tau){V_1(z)}\psi_1(z,\tau)\d z.
\end{equation}
We again assume an isotropic distribution of the recoil momentum $\hbar k$. Contributions with different $\tau$ and $k$
have to be summed in an incoherent way. For the momentum distribution as a function of the wave vector of the
spontaneously emitted photon we get
\begin{equation}
|\phi_k(p)|^2\propto\int_0^{t_{\rm end}}\Gamma(\tau)|\phi_{\tau,k}(p)|^2{\rm d}\tau. \label{phik}
\end{equation}
A subsequent integration over this wave vector yields
\begin{equation}
|\phi(p)|^2=\int_{-k_0}^{k_0}|\phi_k(p)|^2{\rm d}k
\end{equation}
for the momentum distribution of a sample of atoms.

\begin{figure}[t]
\includegraphics[width=\linewidth]{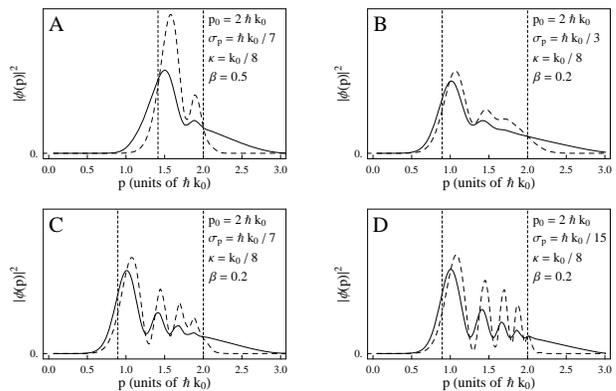}
\caption{Interference patterns calculated for different parameters calculated using the time-dependent method. Solid
lines: $|\phi(p)|^2$, with the effect of a spontaneous recoil, and dashed lines: $|\phi_0(p)|^2$, without the effect of
a spontaneous recoil, versus the final momentum $p$ in units of the photon recoil $\hbar k_0$. The dotted lines
indicate the classically allowed region without recoil. Between (a) and (c) the reduction factor $\beta$ is changed,
(b)-(d) are a sequence where the initial momentum spread $\sigma_p$ is decreased.} \label{fig5}
\end{figure}

Fig. \ref{fig5} shows graphs of $|\phi(p)|^2$ for some parameters. All calculations are performed with $t_{\rm end}=70
m/\hbar k_0^2$ for which the criterium that the entire wave packet has left the potential is fulfilled. Fig.
\ref{fig5}(a) should be compared with Fig. \ref{fig4}(a). The parameters for Figs. \ref{fig5}(b)-(d) are equal to the
parameters used for Fig. \ref{fig4}(c) except for the initial width $\sigma_z$ of the wave packet and thus the momentum
spread $\sigma_p=\hbar/\sigma_z$. Also for a wave packet with a finite momentum spread the interference effects are
present. As expected, the interference fringes are more apparent for a wave packet with a smaller momentum spread.

\section{Experimental considerations}
\label{sec5}

So far we have considered levels $|1\rangle$ and $|2\rangle$ without discussing which physical level they correspond
to. In reality we usually deal with multi-level atoms, that moreover include sub-structure. Each of these (sub-)levels
has a different interaction with the evanescent field. If more (sub-)levels contribute to the signal the predicted
interference can be washed out. For $^{87}$Rb atoms a convenient choice would  be the $|Fm\rangle=|1,0\rangle$ ground
state for state $|1\rangle$ and the $|Fm\rangle=|2,\pm1\rangle$ ground states for state $|2\rangle$. The evanescent
field needs to be linearly ($\pi$) polarized and blue detuned with respect to the $F=1\rightarrow F'=2$ transition of
either the $D_1$ or $D_2$ line. Due to selection rules, only a transition over the $|F'm'\rangle=|2,0\rangle$ excited
state contributes to the transition from state $|1\rangle$ to state $|2\rangle$. This excited state can decay to either
of the $|2,\pm 1\rangle$ ground states by emitting a $\sigma^{\pm}$ polarized photon. Since these states interact
identically with the evanescent field, their interference patterns will overlap.

The necessary linear polarization for the EW can only be easily obtained using a TE ($\perp z$) polarized incident
beam. For a circularly polarized photon this distribution of wave vectors of the spontaneously emitted photon is
non-isotropic. The circularly polarized photon has its quantization axis along the polarization axis of the EW field.
The recoil distribution in the $z$-direction due to such a photon is given by $\frac{3}{16}(3-(k/k_0)^2)$. Intensity
distributions of dipole radiation are given by e.g. \cite{Jackson99}. Since this distribution has a maximum for $k=0$,
for which the recoil-dependent interference patterns are most pronounced as is visible in Fig. \ref{fig3}, the
visibility of the interference signals will be slightly better than presented in this paper.

\section{Discussion and conclusions}

The calculations presented in this paper have been performed for low initial velocities $v_{\rm i}$. This is because
both calculation procedures turned out to be limited by computational resources. For the time-independent approach the
evaluation in {\it Mathematica} \cite{mathematica} of the Bessel-K functions of high imaginary order becomes very slow.
For the time-dependent approach the number of sampling points, necessary for the numerical evaluation, becomes too
large due to the highly oscillatory character of the incident wave packet. However, we expect even better signals for
realistic values for the initial velocity $v_{\rm i}$, so the calculation represents a worst case.

By numerically solving the Schr\"odinger equation we can now reply with an unambiguous yes to the question {\it can the
beamsplitter in an atom interferometer work on the basis of spontaneous emission?} The intuitive objections to whether
this is possible have been refuted. The semi-classical arguments have been confirmed by the full quantum-mechanical
calculations. Which-way information due to the possibility of detecting the time of emission of the spontaneously
emitted photon is avoided by choosing a sufficiently narrow velocity uncertainty. A wave packet covers both transfer
points in phase space simultaneously if its velocity is defined accurately enough. The incoherent nature of a
spontaneous emission process due to the random recoil direction of the atom is visible in all the calculated
interference curves, but does not lead to a complete scrambling of the interference. For larger final velocities, for
which the transfer points are separated more and the acquired random phase is consequently larger, the visibility of
the interference fringes indeed decreases. Furthermore the fringe period indeed qualitatively shows the behavior that
was predicted on the basis of the semi-classical calculations.

Our analysis also shows that the absence or presence of which-way information is not the same as the perturbing effect
of the recoils due to spontaneous transitions. This is most clearly seen in Fig. \ref{fig2} and is in agreement with
the viewpoint of D\"urr {\it et al.} \cite{DurNonRem98}.

\section*{Acknowledgments}

This work is part of the research program of the ``Stichting voor Fundamenteel Onderzoek van de Materie'' (FOM) which
is financially supported by the ``Nederlandse Organisatie voor Wetenschappelijk Onderzoek'' (NWO).

\end{document}